\documentclass[10pt,tightenlines,prl,showpacs,letterpaper,aps,twocolumn,nofootinbib,superscriptaddress]{revtex4-1}

\usepackage{graphicx}
\usepackage{amssymb}
\usepackage{epstopdf}
\usepackage{amsmath,amsthm}
\usepackage[normalem]{ulem} 
\usepackage{comment}
\usepackage{wasysym}
\usepackage{subfigure}
\usepackage{hyperref}
\usepackage{placeins}

\newcommand{\COMMENT}[1]{}

\newcommand{\ket}[1]{\left| #1 \right\rangle}
\newcommand{\bra}[1]{\left\langle #1 \right|}

\newcommand{\beq}{\begin{equation}}
\newcommand{\eeq}{\end{equation}}
\newcommand{\bea}{\begin{align}}
\newcommand{\eea}{\end{align}}
\newcommand{\bq}{\begin{quote}}
\newcommand{\eq}{\end{quote}}

\usepackage{xparse}

\usepackage[x11names]{xcolor}
\ExplSyntaxOn
\newcommand{\colorstring}[3]{%
	\str_set:Nn \l_tmpa_str {#3}
	\int_step_inline:nnnn {1} {1} {\str_count:N \l_tmpa_str } {%
		\int_if_odd:nTF{##1}{
			\textcolor{#1}{\str_item:Nn \l_tmpa_str {##1}}
		}{
			\textcolor{#2}{\str_item:Nn \l_tmpa_str {##1}}
		}%
	}%
}
\ExplSyntaxOff
\begin{document}

\title{
Relationship between covariance of Wigner functions \\ and transformation noncontextuality
}
\author{Lorenzo Catani}\email{lorenzo.catani@tu-berlin.de}\affiliation{Electrical Engineering and Computer Science Department, Technische Universit\"{a}t Berlin, 10587 Berlin, Germany}

\begin{abstract}
We investigate the relationship between two properties of quantum transformations often studied in popular subtheories of quantum theory: covariance of the Wigner representation of the theory and the existence of a transformation noncontextual ontological model of the theory. 
We consider subtheories of quantum theory specified by a set of states, measurements and transformations, defined specifying a group of unitaries, that map between states (and measurements) within the subtheory. We show that if there exists a Wigner representation of the subtheory which is covariant under the group of unitaries defining the set of transformations then the subtheory admits of a transformation noncontextual ontological model. 
We provide some concrete arguments to conjecture that the converse statement also holds provided that the underlying ontological model is the one given by the Wigner representation.  
In addition, we investigate the relationships of covariance and transformation noncontextuality with the existence of a quasiprobability distribution for the theory that represents the transformations as positivity preserving maps. We conclude that covariance implies transformation noncontextuality, which implies positivity preservation.
\end{abstract}

\maketitle

\section{Introduction}
Understanding what are the features of quantum theory that resist any explanation within the classical worldview is crucial both from the foundational and the computational point of view. Several no-go theorems have provided concrete contributions in this sense \cite{Bell1966,KochenSpecker67,Leifer2017,PBR,FrauchigerRenner2018,CataniLeifer2020}, as well as results concerning attempts of reproducing quantum theory in a classical fashion \cite{Wigner1932,Spekkens2007,ToyFieldTheory} and axiomatizations of quantum theory in the framework of general probabilistic theories \cite{Hardy2001, Chiribella2011, Masanes2011}. 
Among the notions of nonclassicality, in recent years contextuality \cite{KochenSpecker67, Spekkens2005} has also been employed to explain the origin of the quantum computational speed-up for specific tasks and models of quantum computation \cite{Howard2014, Raussendorf2017, Delfosse2015, Vega2017, CataniBrowne2018, Raussendorf2013, Spekkens2009,Mansfield2018,CataniHenaut2018, Raussendorf2019,Schmid2018,Saha2019,SahaAnubhav2019,LostaglioSenno2020,Yadavalli2020,ContextualityViaUR,Flatt2021,Roch2021,CataniFaleiro2022}.
Nevertheless, there is still a neat discrepancy between what is considered to be truly nonclassical from a foundational and a computational point of view. An example is provided by the $n-$qubit stabilizer theory -- the subtheory of quantum theory composed by common eigenstates of Pauli operators, Clifford unitaries and Pauli measurements -- which has been proven to be efficiently simulatable by a classical computer according to Gottesman-Knill theorem \cite{Gottesman99}, despite it shows contextuality \cite{Mermin1990, GHZ}. 
It would be desirable to obtain a consistent picture connecting the notions of nonclassicality in the two realms and distinguishing weaker and stronger notions of nonclassicality. In this respect, a fruitful approach would be to explore less studied and new notions of (non)classicality.
 
In the present work we move a step in this direction. We start by studying two properties of quantum transformations that are usually associated to classical behaviors and therefore, when broken, to genuinely nonclassical features. 
These are the covariance of the Wigner function under the group of transformations allowed in the examined theory \cite{Zhu2016}, and transformation noncontextuality -- \textit{i.e.} the existence of a transformation noncontextual ontological model for the theory \cite{Spekkens2005}. 
Let us spend few words to introduce them and to discuss why they should be considered as notions of classicality.

Covariance is defined in the framework of Wigner functions \cite{Wigner1932} -- particular quasiprobability distributions that provide a representation of quantum states, transformations and measurements in the phase space -- and it indicates that the transformations of the theory under examination can be represented as symplectic affine transformations in the phase space (a subset of the permutations in the discrete dimensional case) \cite{Gross2006,Zhu2016}.
It is often studied in reference to the group of Clifford unitaries, and it is known that a covariant representation of such group exists in the odd dimensional case \cite{Gross2006}, but it does not exist in the even dimensional case \cite{Raussendorf2022}. This is why, in the even dimensional case, smaller subtheories than the stabilizer theory have to be considered in order to have covariance \cite{Raussendorf2017,Delfosse2015}.
In the context of studying quantum theory in the phase space, if one wonders what it means for a set of quantum transformations to have a classical behavior, it is natural -- at least as a minimal requirement -- to assume this to be the case if the quantum transformations are represented in the phase space in the same way as the physical trajectories in classical Hamiltonian mechanics. This means, by symplectic transformations (necessary and sufficient condition, as proven by Hamilton \cite{Guillemin1990}). Hence, covariantly represented quantum transformations can be interpreted as classical trajectories in the phase space. In addition, covariance turns out to capture a wider notion of classicality than the mere correspondence to trajectories in classical Hamiltonian mechanics.  The set of quantum transformations that are reproduced in the celebrated Spekkens' toy theory -- a noncontextual theory formulated in the phase space that reproduces many phenomena that are usually thought to be signatures of quantumness \cite{Spekkens2007, Spekkens2016, CataniBrowne2017,Catani2021} -- allows for a covariant Wigner representation. The reason is that covariance of the Wigner representation coincides with the requirement of preservation of the ``epistemic restriction'' (corresponding to the preservation of the Poisson brackets), which is the basic defining constraint of Spekkens toy theory. 
Finally, a consequence of the present work is that covariance is indeed a notion of classicality also because it implies transformation noncontextuality, which is a well established notion of classicality, as we discuss below.

Transformation noncontextuality is defined in the framework of ontological models (also known as hidden variable models) \cite{Harrigan2010}. As we will report in the next section, quasiprobability representations and ontological models of a theory are strictly related notions \cite{Spekkens2008}. 
Transformation noncontextuality means that operationally equivalent experimental procedures for performing a transformation must correspond to the same representation in the underlying ontological model \cite{Spekkens2005}. It is often justified as an instance of a methodological principle motivated by Leibniz's principle of the identity of indiscernibles \cite{SpekkensLeibniz2019}, and more recently as a requirement of no fine tuning \cite{CataniLeifer2020,Adlam2021}. 
Transformation noncontextuality holds in classical mechanics. However, it was proven in \cite{Spekkens2008} that it does not exist a transformation noncontextual ontological model consistent with the statistics of quantum theory.

Historically, most research on the role of nonclassical features in quantum computation has focused on properties of preparations and measurements \cite{WallmanBartlett,Howard2014, Raussendorf2017}. Instead, we here study properties of quantum transformations. They have recently been shown to play a crucial role in information processing tasks \cite{Mansfield2018,CataniHenaut2018}, and in encoding nonclassical behaviors of subtheories of quantum theory that were previously considered to behave classically, like the single qubit stabilizer theory \cite{Lillystone2018}. 
The latter provides one of the main motivations for considering covariance and transformation noncontextuality as connected: they are both broken by the presence of the Hadamard gate.
Another important result that ignores transformations is due to Spekkens in \cite{Spekkens2008}, where he showed that contextuality and negativity of quasiprobability representations are equivalent notions of nonclassicality. 

In this article we focus on subtheories of quantum theory defined by a set $(\mathcal{S},\mathcal{T},\mathcal{M})$ of states, transformations and measurements, where the transformations map between states (and measurements) within the subtheory. Because covariance only applies to unitaries (it would not make sense for other transformations, \textit{e.g.}, those describing irreversible processes), we only consider theories whose set of transformations $\mathcal{T}$ is defined by a group of unitaries, and where the more general channels can be decomposed as convex mixtures of unitaries. 

By using the result in \cite{Spekkens2008}, that trivially extends to transformations, we show that covariance implies transformation noncontextuality (proposition \ref{CovarianceTransfNC}). We conjecture that the converse holds with the assumption that the ontological model is the one corresponding to the Wigner representation. We provide some concrete arguments to support such claim in conjecture \ref{TransfNCCovariance}. 

In this work we also introduce the notion of positivity preservation. We do so with the goal to potentially formulate a stronger notion of nonclassicality than transformation contextuality and non-covariance, that are quite weak insofar as they manifest also in simple classically simulatable subtheories like the single qubit stabilizer theory. 
We say that a subtheory is positivity preserving if there exists a quasiprobability representation for which all the transformations of the theory map non-negative states to non-negative states. 
Positivity preservation 
 holds in classical mechanics, where quasiprobabilities are (non-negative) probabilities, and it does not hold in full quantum theory, because of the unavoidable presence of negatively represented states and the fact that there always exists a transformation between any two states. 
 
In propositions \ref{TheoremCovariance} and \ref{TheoremTransfNC} we prove that both covariance and transformation noncontextuality imply positivity preservation, but not vice versa. 
These results motivate further research on the conceptual significance of positivity preservation as a compelling notion of classicality, since it would then be the case, via propositions \ref{TheoremCovariance} and \ref{TheoremTransfNC}, that breaking positivity preservation is a stronger notion of nonclassicality than transformation contextuality and non-covariance.

The remainder of the article is structured as follows. In section \ref{Definitions} we define quasiprobability distributions and Wigner functions  (subsection \ref{SecWignerFunctions}), covariance of Wigner functions (subsection \ref{SecCovariance}), and transformation noncontextuality  (subsection \ref{SecTransformationNoncontextuality}).  In section \ref{SecSQM} we illustrate the motivating example of the single qubit stabilizer theory and then, in section \ref{Results}, we prove the results relating covariance, transformation noncontextuality and positivity preservation. We conclude, in section \ref{Discussion}, by discussing possible generalizations of the results and future avenues.


\section{Definitions}
\label{Definitions}
In this section we define Wigner functions and more general quasiprobability distributions (with a special focus on the property of positivity preservation), covariance of Wigner functions, and transformation noncontextuality.

\subsection{Wigner functions} 
\label{SecWignerFunctions}
Wigner functions are a way of reformulating quantum theory in the phase space \cite{Wigner1932}, which is the framework where classical Hamiltonian mechanics is formulated, and therefore provide a tool to compare the two theories on the same ground. They are also the most popular example of quasiprobability distributions \cite{Optics, Gross2006, Raussendorf2017}. The latter are linear and invertible maps from operators on Hilbert space to real distributions on a measurable space $\Lambda.$ Quasiprobability representations of quantum theory associate \textit{i)} a real-valued function $\mu_{\rho}: \Lambda \rightarrow \mathbb{R}$ to any density operator $\rho$ such that $\int d\lambda \mu_{\rho}(\lambda)=1$, \textit{ii)} a real-valued function $\xi_{\Pi_k}: \Lambda \rightarrow \mathbb{R}$ to any element $\Pi_k$ of the POVM $\{\Pi_k\}$ such that $\sum_k \xi_{\Pi_k}(\lambda)=1$ $\;\forall \; \lambda\in\Lambda,$ and \textit{iii)} a real-valued matrix $\Gamma_{\varepsilon}: \Lambda\times\Lambda \rightarrow \mathbb{R}$ to any CPTP map $\varepsilon,$ such that $\int d\lambda' \Gamma_{\varepsilon}(\lambda',\lambda)=1.$ These distributions provide the same statistics of quantum theory -- given by the Born rule -- if \begin{equation}\begin{split} \label{QuantumStatistics} p(k|\rho,\varepsilon, \{\Pi_k\}) &= \textrm{Tr}[\varepsilon(\rho)\Pi_k] \\ &= \int d\lambda d\lambda' \xi_{\Pi_k}(\lambda') \Gamma_{\varepsilon}(\lambda',\lambda) \mu_{\rho}(\lambda).\end{split}\end{equation}
Quasiprobability distributions are named this way as they behave similarly to probability distributions, with the crucial difference that they can take negative values. The negativity is usually assumed to be a signature of quantumness because it is unavoidable in order to reproduce the statistics of quantum theory \cite{FerrieEmerson2009}.

A quasiprobability distribution, for example $\mu_{\rho}(\lambda),$ is \textit{non-negative} if $\mu_{\rho}(\lambda)\ge0 \;\; \forall \lambda\in\Lambda.$ Non-negative quasiprobability representations of quantum theory are an important tool to support the epistemic view of quantum theory \cite{Harrigan2010,Spekkens2007}, where quantum states are interpreted as states of knowledge rather than states of reality, and are useful to perform classical simulations of quantum computations \cite{Veitch2012}. 
We will focus on the following property of quasiprobability distributions.

\newtheorem{Definition}{Definition}
\begin{Definition}[Positivity Preservation]\label{PositivityPreservation}
Given a set $\mathcal{S}_+$ of quantum states $\rho$ that are \emph{non-negatively} represented by a quasiprobability distribution $\mu_{\rho}(\lambda),$ a transformation $\varepsilon$ that maps $\rho$ to $\rho'=\varepsilon(\rho)$ is \emph{positivity preserving} if, for every $\rho\in\mathcal{S}_+,$  the quasiprobability distribution $\mu_{\rho'}(\lambda)$ is non-negative too. 

A subtheory of quantum theory allows for a positivity preserving quasiprobability representation if there exists a quasiprobability distribution for which all the transformations of the subtheory are positivity preserving.
\end{Definition}

We are interested in Wigner functions defined over the discrete phase space $\Lambda=\mathbb{Z}^{2n}_{d},$ where the integers $d,n$ denote the dimensionality of the system and the number of systems, respectively. We follow the definition provided in \cite{Raussendorf2017} that includes the mostly used examples of Wigner functions \cite{Gross2006,Gibbons2004}. 

\newtheorem{Wigner}[Definition]{Definition}
\begin{Wigner}[Wigner function]\label{WignerFunctions}
The Wigner function of a quantum state $\rho$ (and, analogously, of a POVM element $\Pi_k$) is defined as
\begin{equation}\label{wignernew}W^{\gamma}_{\rho}(\lambda)=\textrm{Tr}[A^{\gamma}(\lambda)\rho],\end{equation} where the phase-point operator is \begin{equation}\label{phasepoint} A^{\gamma}(\lambda)= \frac{1}{N_{\Lambda}}\sum_{\lambda' \in\Lambda}\chi([\lambda,\lambda'])\hat{W}^{\gamma}(\lambda').\end{equation}
\end{Wigner}
The normalization $N_{\Lambda}$ is such that $\textrm{Tr}(A^{\gamma}(\lambda))=1.$
The Weyl operators $\hat{W}^{\gamma}(\lambda)$ are defined as  $\hat{W}^{\gamma}(\lambda)=w^{\gamma(\lambda)}Z(p)X(x),$ where the phase space point is $\lambda=(x,p)\in \Lambda.$ 
The operators $X(x),Z(p)$ represent the (generalized) Pauli operators, $X(x)=\sum_{x'\in\mathbb{Z}_d} \left | x'-x \right \rangle \left \langle x' \right |$ and $Z(p)= \sum_{x\in\mathbb{Z}_d} \chi(px)\left | x \right \rangle \left \langle x \right |.$
The functions $w$ and $\chi$ are $w^{\gamma(\lambda)}=i^{\gamma(\lambda)}, \chi(a)=(-1)^a$ in the case of qubits and $w^{\gamma(\lambda)}=\chi(-2^{-1}\gamma(\lambda)),\chi(a)=e^{\frac{2\pi i}{d}a}$ in the case of qudits of odd dimensions.
By choosing the function $\gamma: \Lambda\rightarrow \mathbb{Z}_q,$ where $q$ is an integer,  
we can specify which Wigner function to consider. 

The reason why choosing different Wigner functions is that they are non-negative -- thus providing a proof of operational equivalence with classical theory -- for different subtheories of quantum theory. 
For example $\gamma= x\cdot p$ gives Gross' Wigner function \cite{Gross2006,Gross2019} for odd dimensional qudits, which non-negatively represents stabilizer quantum theory in odd dimensions \cite{Spekkens2016,CataniBrowne2017,SchmidGrossUnique2022}, while $\gamma=x \cdot p \;\mathrm{mod}4$ gives the Wigner function developed by Gibbons and Wootters \cite{Gibbons2004,Cormick2006} for qubits, which non-negatively represents the subtheory of quantum theory composed by all the separable eigenstates of Pauli operators, Pauli measurements and transformations between them \cite{Raussendorf2017,CataniBrowne2018}. 
We will omit the superscript $\gamma$ in the future in order to soften the notation. 

Wigner functions satisfy the following properties.
\begin{itemize}
\item The marginal of a Wigner function on the state $\rho$ behaves as a probability distribution: $\sum_{p\in \mathbb{Z}_d}W_{\rho}(x,p)=|\bra{x}\rho\ket{x}|^{2}.$
\item The Wigner function of many systems in a product state is the tensor product of the Wigner functions of each system, as a consequence of the factorability of $A(\lambda),$ $A(x_0,p_0\dots x_{n-1},p_{n-1})=A(x_0,p_0)\otimes \dots \otimes A(x_{n-1},p_{n-1}).$ 
\item The phase-point operators form a complete basis of the Hermitian operators in the Hilbert space with respect to the Hilbert-Schmidt inner product, thus obeying Hermitianity, $A(\lambda)=A^{\dagger}(\lambda)$ $\forall \; \lambda\in\Lambda,$ and orthonormality, $\textrm{Tr}[A(\lambda)A(\lambda')]=\frac{1}{N_{\Lambda}}\delta_{\lambda,\lambda'}.$
This implies that $\rho=\sum_{\lambda\in\Lambda}A(\lambda)W_{\rho}(\lambda)$ and $\sum_{\lambda\in\Lambda}W_{\rho}(\lambda)W_{\sigma}(\lambda)=\textrm{tr}(\rho\sigma),$ where $\rho,\sigma$ are any two Hermitian operators.
\item $\sum_{\lambda\in\Lambda}A(\lambda)=\mathbb{I},$ thus implying that $\textrm{Tr}[\rho]=1=\sum_{\lambda}W_{\rho}(\lambda).$
\end{itemize}
For consistency with equation \eqref{QuantumStatistics}, the Wigner function associated to a CPTP map $\varepsilon$ is $W_{\varepsilon}(\lambda',\lambda)=\textrm{Tr}[\varepsilon(A(\lambda))A(\lambda)]$ and it is such that $\sum_{\lambda\in\Lambda}W_{\varepsilon}(\lambda',\lambda)=1.$



\subsection{Covariance}
\label{SecCovariance}

The Wigner function $W_{\rho}(\lambda)$ is \textit{covariant} under the group\footnote{Notice that the set of unitaries must form a group for the property of covariance to be possible.} of unitary transformations in $\mathcal{T}$ if, for every unitary gate $U\in\mathcal{T}$, 
\begin{equation}\label{covariance}W_{U\rho U^{\dagger}}(\lambda)=W_{\rho}(S\lambda+a)  \;\;\; \forall \;\rho\in\mathcal{S},\; \forall \;\lambda\in\Lambda,\end{equation} where $S$ is a symplectic transformation, \textit{i.e.} $S^TJS=J,$ where $J= \bigoplus_{j=1}^{n} \begin{bmatrix} 0 & 1 \\ -1 & 0 \end{bmatrix}_j$ is the standard invertible matrix used in symplectic geometry, and $a$ is a translation vector. There is a precise relationship between the group of unitaries and the group of symplectic transformations satisfying equation \eqref{covariance}; namely, the unitaries $U$ are indexed by the symplectic transformations $S$, and form a unitary representation of the group of such symplectic transformations.

We say that the set of transformations $\mathcal{T}$, as well as the subtheory $(\mathcal{S},\mathcal{T},\mathcal{M})$, is covariantly represented by the Wigner function $W_{\rho}(\lambda)$ if $W_{\rho}(\lambda)$ is covariant under the group of unitary transformations in $\mathcal{T}$ according to equation \eqref{covariance}. 
Equation \eqref{covariance} can also be written as, \begin{equation}\label{covarianceA}UA(\lambda)U^{\dagger}=A(S\lambda+a)\;\;\; \forall \;\lambda\in\Lambda.\end{equation}
In the subtheories that we consider, the allowed quantum channels are the ones decomposed as mixtures of unitaries in $\mathcal{T}$. We recall that a quantum channel $\varepsilon$ -- represented by a CPTP map  -- can in general be written as \begin{equation}\label{KrausDecomposition}\varepsilon(\rho)=\sum_kE_k\rho E_k^{\dagger},\end{equation} where the Kraus operators $E_k$ satisfy the completeness relation $\sum_kE_k^{\dagger}E_k=\mathbb{I}.$ 
In our case, the Kraus operators are of the form $E_k = \sqrt{p_k} U_k$, where $U_k$ are unitaries in $\mathcal{T}$ and the $p_k\in[0,1]$ are such that $\sum_k p_k =1$. 
We say that a channel $\varepsilon$ is covariantly represented by the Wigner function $W_{\rho}(\lambda)$ if it belongs to the set of transformations  $\mathcal{T}$ that is covariantly represented by the Wigner function $W_{\rho}(\lambda)$.



\subsection{Transformation Noncontextuality}
\label{SecTransformationNoncontextuality}
A natural way of justifying why quantum theory works is to provide an ontological model that reproduces its statistics \cite{Spekkens2005}. An ontological model associates the physical state of the system at a given time -- the ontic state -- to a point $\lambda$ in a measurable set $\Lambda,$ and the experimental procedures -- classified in preparations, transformations and measurements -- to probability distributions on the ontic space $\Lambda.$ A preparation procedure $P$ of a quantum state $\rho$ is represented by a probability distribution $\mu_P(\lambda)$ over the ontic space, $\mu_P:\Lambda\rightarrow \mathbb{R}$ such that $\int \mu_P(\lambda)d\lambda=1$ and $\mu_{P}(\lambda)\ge0 \;\; \forall \lambda\in\Lambda.$ A transformation procedure $T$ of a CPTP map $\varepsilon$ is represented by a transition matrix $\Gamma_T(\lambda',\lambda)$ over the ontic space, $\Gamma_T:\Lambda\times\Lambda\rightarrow \mathbb{R}$ such that $\int \Gamma_T(\lambda',\lambda)d\lambda'=1$ and $\Gamma_{T}(\lambda',\lambda)\ge0 \;\; \forall \lambda,\lambda'\in\Lambda.$ A measurement procedure $M$ with associated outcomes $k$ of a POVM $\{\Pi_k\}$ is represented by a set of indicator functions $\{\xi_{M,k}(\lambda)\}$ over the ontic space, $\xi_{M,k}:\Lambda\rightarrow \mathbb{R}$ such that $\sum_k \xi_{M,k}(\lambda)=1$ and $\xi_{M,k}(\lambda)\ge0 \;\; \forall \lambda\in\Lambda,\;\forall k.$ The ontological model reproduces the predictions of quantum theory according to the law of classical total probability, \begin{equation}\begin{split} \label{QuantumStatisticsOM} p(k|P,T,M) &= \textrm{Tr}[\varepsilon(\rho)\Pi_k] \\ &= \int d\lambda d\lambda' \xi_{M,k}(\lambda') \Gamma_T(\lambda',\lambda) \mu_P(\lambda).\end{split}\end{equation}
All the preparation procedures that prepare the state $\rho$ belong to the equivalence class that we denote with $e_{\rho}(P).$ Analogous reasoning for the equivalence classes $e_{\varepsilon}(T)$ associated to the CPTP map $\varepsilon$ and $e_{\{\Pi_k\}}(M)$ associated to the POVM $\{\Pi_k\}.$ 
The idea of the generalized notion of noncontextuality is that operational equivalences -- \textit{e.g.}, the different Kraus decompositions of a CPTP map -- are represented by identical probability distributions on the ontic space.
\newtheorem{TransformationNC}[Definition]{Definition}
\begin{TransformationNC}[Ontological model]\label{TransformationNC}
An ontological model of (a subtheory of) quantum theory is \emph{transformation noncontextual} if \begin{equation}\label{TransfNC}\Gamma_T(\lambda',\lambda)=\Gamma_{\varepsilon}(\lambda',\lambda) \;\;\; \forall \;T\in e_{\varepsilon}(T),  \;\; \forall \;\varepsilon.\end{equation}
\end{TransformationNC}
The above definition can be analogously extended to preparation noncontextuality and measurement noncontextuality \cite{Spekkens2005}, \begin{equation}\begin{split}\label{PrepMeasNC}&\mu_P(\lambda)=\mu_{e_{\rho}(P)}(\lambda) \;\;\;\forall P\in e_{\rho}(P), \;\; \forall \;\rho, \\ & \xi_{M,k}(\lambda)=\xi_{e_{\Pi_k}(M)}(\lambda) \;\;\; \forall M \in e_{\{\Pi_k\}}(M), \;\; \forall \;\{\Pi_k\}.\end{split}\end{equation} 
An ontological model is universally noncontextual if it is preparation, transformation and measurement noncontextual.
It can be shown that a universally noncontextual ontological model of quantum theory is impossible \cite{Spekkens2005} and, in particular, that a transformation noncontextual ontological model of quantum theory is impossible too.

From the definitions of ontological models above and the definitions of quasiprobability representations provided previously, by substituting equations \eqref{TransfNC} and \eqref{PrepMeasNC} into equation \eqref{QuantumStatisticsOM}, it is immediate to see that the existence of a \textit{non-negative} quasiprobability representation that provides the statistics of quantum theory as in equation \eqref{QuantumStatistics} coincides with the existence of a \textit{noncontextual} ontological model for it. This result was proven in \cite{Spekkens2008} and here stated also considering transformations.


\section{The single qubit stabilizer theory}
\label{SecSQM}
In this section we consider the stabilizer theory of one qubit, which provides a motivating example for studying the connection between covariance and transformation noncontextuality, as we will now show. 

The stabilizer theory of one qubit is defined as the subtheory of one qubit quantum theory that includes the eigenstates of $X,Y,Z$ Pauli operators, the Clifford unitaries -- generated by the Hadamard gate $H$ and the phase gate $P$ -- and $X,Y,Z$ Pauli observables.
Its states and measurement elements are non-negatively represented by Wootters-Gibbons' Wigner functions \cite{Gibbons2004}, and the transformations are positivity preserving. 
More precisely, as shown in \cite{Cormick2006}, there are two possible Wigner functions for one qubit stabilizer states that are non-negative and obey the desired properties stated in the previous section. 
We denote the two Wigner functions with $W_{+}$ and $W_{-}.$ The corresponding phase-point operators are \[A_+(0,0)=\frac{1}{2}(\mathbb{I}+X+Y+Z),\] \[A_-(0,0)=\frac{1}{2}(\mathbb{I}+X+Y-Z),\] and $A_{j}(0,1)=XA_{j}(0,0)X^{\dagger},$  $A_{j}(0,1)=YA_{j}(0,0)Y^{\dagger},$ $A_{j}(1,1)=ZA_{j}(0,0)Z^{\dagger}$ for $j\in\{+,-\},$ where we have denoted with $X,Y,Z$ both the Pauli observables and the Pauli transformations. 
The Wigner functions above are \textit{not} covariant under the group of Clifford unitaries. The reason is that the Hadamard gate (or, similarly, the phase gate) unavoidably maps a phase-point operator to a phase-point operator belonging to a different basis set, \textit{e.g.}, $HA_+(0,0)H=A_-(0,1),$ as a consequence of its action on the Pauli operators, $HXH=Z, HZH=X$ and $HYH=-Y.$ 

\begin{figure}[h!]
\centering
{\includegraphics[width=.4\textwidth,height=.22\textheight]{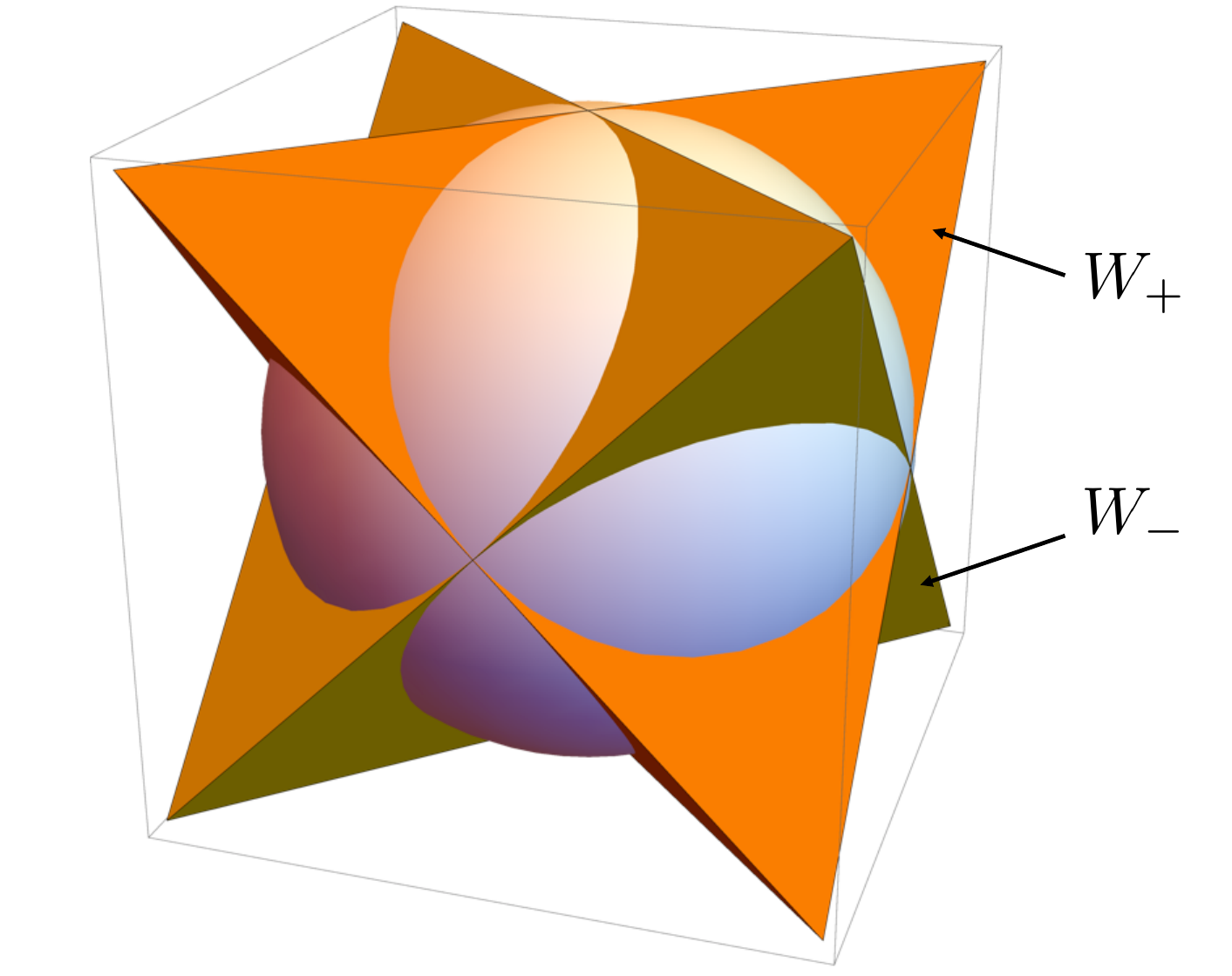}}
\caption{\textbf{Non-negative Wigner functions of one qubit stabilizer states.} The figure above shows the states that are non-negative according to the two Wigner functions $W_{+}$ and $W_{-}$ for one qubit defined in \cite{Cormick2006}. The intersection of the two non-negative sectors in the Bloch sphere forms an octahedron, which is the state space of the stabilizer theory of one qubit. The idea behind the 8-state model \cite{WallmanBartlett} is to consider a model whose ontic space has twice the dimension of the standard phase space and consider both the two Wigner functions. 
The Hadamard gate, as well as the phase gate, maps between the two Wigner functions, thus causing the breaking of covariance and the distinction between operationally equivalent realizations of the completely depolarizing channel, \textit{i.e.} the presence of transformation contextuality.}
\label{BlochWF}
\end{figure}

A popular preparation and measurement noncontextual ontological model of the single qubit stabilizer theory is the 8-state model \cite{WallmanBartlett}. The quantum states (and measurement elements) are represented as uniform probability distributions over an ontic space of dimension $8$ (twice the dimension of the standard phase space) and the Clifford transformations are represented by permutations over the ontic space. Intuitively, it can be seen as a model that corresponds to take into account both the Wigner functions $W_{+}$ and $W_{-},$ as shown in figure \ref{BlochWF}.\footnote{Notice that the 8-state model does \textit{not} correspond to a quasiprobability distribution as defined in the previous section.}

The 8-state model may be viewed as another confirmation of the classical nature of the theory. However, it was proven in \cite{Lillystone2018} that this model and, more in general, any preparation and measurement noncontextual model of the single qubit stabilizer theory, shows transformation contextuality. This is due again, as for the breaking of covariance, to the Hadamard gate (or, similarly, to the phase gate). More precisely, if we consider the operationally equivalent decompositions of the completely depolarizing channel, $\varepsilon_1(\rho)=\sum_RR\rho R=\mathbb{I}$ and $\varepsilon_2(\rho)=\sum_R(HR)\rho (HR)^{\dagger}=\mathbb{I},$ where $R\in\{\mathbb{I},X,Y,Z\}$ and $\rho$ is any qubit stabilizer state, they are ontologically distinct in any preparation and measurement noncontextual ontological model of the single qubit stabilizer theory \cite{Lillystone2018}. 

The ontological distinctness can be pictured in the 8-state model, where $\varepsilon_1$ and $\varepsilon_2$ are represented as different permutations in the ontic space. The former only maps within a half of the ontic space, that can be seen as the phase space where $W_+$ (or $W_-$) is defined, while the latter maps between the two halves of the ontic space, so between the phase space of $W_+$ and the phase space of $W_-$, which is also a way of picturing the breaking of covariance. This last consideration suggests the presence of a tight connection between covariance and transformation noncontextuality.


\section{Results}
\label{Results}
We now relate the properties of quantum transformations defined so far.
\newtheorem{Proposition}{Proposition}
\begin{Proposition} \label{CovarianceTransfNC} Let us consider a subtheory of quantum theory $(\mathcal{S},\mathcal{T},\mathcal{M})$, where the set $\mathcal{T}$ is made of transformations as defined in equation~\eqref{KrausDecomposition}.
If there exists a Wigner function by which the subtheory is covariantly represented, then the subtheory admits of a transformation noncontextual ontological model.
\end{Proposition}

\begin{proof}
Let us consider any transformation $\varepsilon\in\mathcal{T}$, with Kraus decomposition 
$\varepsilon(\rho)=\sum_kE_k\rho E_k^{\dagger}$ for every $\rho\in\mathcal{S},$ where $E_k = \sqrt{p_k} U_k$, $U_k$ are unitaries in $\mathcal{T}$ and $p_k\in[0,1]$ are such that $\sum_k p_k =1$.

Given covariance (equation~\eqref{covarianceA}), the orthonormality of the phase-point operators, and the linearity of the trace, then $W_{\varepsilon}(\lambda',\lambda)=\textrm{Tr}[\varepsilon(A(\lambda))A(\lambda')]=\textrm{Tr}[\sum_kA(S_k\lambda+a_k)A(\lambda')]=1/N_{\Lambda}\sum_k\delta_{S_k\lambda+a_k,\lambda'}\ge0.$ A non-negative Wigner function for the transformations $\varepsilon\in\mathcal{T}$ implies a transformation noncontextual ontological model for the subtheory, as we have shown at the end of subsection \ref{SecTransformationNoncontextuality}. 

\end{proof}

In short, proposition \ref{CovarianceTransfNC} states that, given our definition of subtheories, covariance implies transformation noncontextuality. However, the converse implication may not hold in general. In particular, it could be possible to have a transformation noncontextual ontological model, even if the ontological model corresponding to the Wigner representation is transformation contextual.\footnote{We leave the search for such a counter example -- i.e. of a subtheory that admits of a transformation noncontextual ontological model but does not admit of a covariant Wigner function -- as future research.} For this reason, the more sensible question to ask is: restricting to the cases where the ontological model corresponds to a Wigner representation, is it true that transformation noncontextuality implies covariance?

We conjecture this to be the case. We support this conjecture by showing that the existence of a transformation noncontextual ontological model corresponding to a Wigner representation of the subtheory $(\mathcal{S},\mathcal{T},\mathcal{M})$ implies that phase point operators are mapped between themselves by the unitary transformations in $\mathcal{T}$. Notice that this does not constitute a proof of the conjecture because having phase point operators being mapped between themselves by the unitary transformations of the theory does not say anything about such maps being symplectic, and hence covariance is not implied. 

\newtheorem{Conjecture}{Conjecture}
\begin{Conjecture} \label{TransfNCCovariance} Let us consider a subtheory of quantum theory $(\mathcal{S},\mathcal{T},\mathcal{M})$, where the set $\mathcal{T}$ is made of transformations as defined in equation~\eqref{KrausDecomposition}.
If there exists a Wigner representation that provides a transformation noncontextual ontological model for the subtheory then the subtheory is covariantly represented by such Wigner function.
\end{Conjecture}

\begin{proof}[Proof of the argument in support of the conjecture]
We here show that the existence of a transformation noncontextual ontological model corresponding to a Wigner representation of the subtheory $(\mathcal{S},\mathcal{T},\mathcal{M})$ implies that phase point operators are mapped between themselves by the unitary transformations in $\mathcal{T}$.

Let us consider a Wigner representation that provides a noncontextual ontological model of the subtheory. This means that the Wigner functions $ W_\varepsilon (\lambda|\lambda’) = \textrm{Tr}[\varepsilon(A(\lambda))A(\lambda')]$ are non-negative for every $\varepsilon\in\mathcal{T}$, where $\varepsilon(\rho)=\sum_kE_k\rho E_k^{\dagger}$ for every $\rho\in\mathcal{S},$  $E_k = \sqrt{p_k} U_k$, $U_k$ are unitaries in $\mathcal{T}$ and $p_k\in[0,1]$ are such that $\sum_k p_k =1$. In particular, this means that also the special case where $\varepsilon$ corresponds to one of the unitaries $U\in\mathcal{T}$ is represented non-negatively, \textit{i.e.}, $W_{U} (\lambda|\lambda’)=\textrm{Tr}[UA(\lambda)U^{\dagger}A(\lambda')]\ge0.$


Let us now define $B\equiv U A(\lambda)U^{\dagger}$ and, by the fact that the phase-point operators $A(\lambda)$ form a basis for the Hermitian operators in the Hilbert space, let us write it as $B=\sum_{\lambda'}W_{U}(\lambda|\lambda')A(\lambda').$ Successively, notice, from $\textrm{Tr}[A(\lambda)]=1,$ that  $\textrm{Tr}[B]=\sum_{\lambda'}W_{U}(\lambda | \lambda')=1.$ Moreover, from the orthonormality of the phase-point operators, $\textrm{Tr}[B^2]= \sum_{\lambda'}W^2_{U}(\lambda | \lambda')=1.$ 
Therefore, given that $W_{U}(\lambda|\lambda')\ge0,\sum_{\lambda'}W_{U}(\lambda | \lambda')=1,$ and $\sum_{\lambda'}W^{2}_{U}(\lambda | \lambda')=1,$ we conclude that $W_{U}(\lambda|\lambda')=0$ $\forall \lambda'$ apart from one $\tilde{\lambda'},$ for which $W_{U}(\lambda|\tilde{\lambda'})=1.$ This means that $B$ coincides with one of the phase-point operators, \textit{i.e.} $U$ maps a phase point operator into another phase point operator. This is true for every $U\in\mathcal{T}$. 

\end{proof}

Let us now state the relationships of covariance and transformation noncontextuality with the existence of a positivity preserving quasiprobability distribution for the subtheory. 

\newtheorem{TheoremCovariance}[Proposition]{Proposition}
\begin{TheoremCovariance}\label{TheoremCovariance} Let us consider a subtheory of quantum theory $(\mathcal{S},\mathcal{T},\mathcal{M})$, where the set $\mathcal{T}$ is made of transformations as defined in equation~\eqref{KrausDecomposition}. If there exists a Wigner function by which the subtheory is covariantly represented, then the subtheory allows for a positivity preserving quasiprobability representation. The converse implication does \emph{not} hold, \textit{i.e.}, if the subtheory allows for a positivity preserving quasiprobability representation, then it is not always the case that there exists a Wigner function by which the subtheory is covariantly represented.
\end{TheoremCovariance}

\begin{proof}
Let us consider the states $\rho\in\mathcal{S}$ with non-negative Wigner function $W_{\rho}(\lambda)$ and any covariantly represented transformation $\varepsilon$ which maps $\rho$ to $\rho'=\varepsilon(\rho)$ such that $W_{\rho'}(\lambda)=\sum_k W_{\rho}(S_k\lambda+a_k).$ Since all $W_{\rho}(S_k\lambda+a_k)$ are non-negative because $W_{\rho}(\lambda)\ge0$ $\forall \lambda\in\Lambda$, then also $W_{\rho'}(\lambda)$ is non-negative. 
The converse implication does not hold, as proven by the counterexample of the single qubit stabilizer theory, where the Hadamard gate maps non-negative states to non-negative states (Wootters-Gibbons' Wigner function \cite{Gibbons2004,Cormick2006}), but it is not covariantly represented.  
\end{proof}

\newtheorem{TheoremTransfNC}[Proposition]{Proposition}
\begin{TheoremTransfNC}\label{TheoremTransfNC} Let us consider a subtheory of quantum theory $(\mathcal{S},\mathcal{T},\mathcal{M}).$ If the subtheory allows for a transformation noncontextual ontological model, then it also allows for a positivity preserving quasiprobability representation. The converse implication does \emph{not} hold, \textit{i.e.}, if the subtheory allows for a positivity preserving quasiprobability representation, then it is not always the case that the subtheory admits of a transformation noncontextual ontological model.

\end{TheoremTransfNC}

\begin{proof}
From Spekkens' result \cite{Spekkens2008} extended to transformations, the existence of a transformation noncontextual ontological model for the subtheory implies the existence of a quasiprobability representation that non-negatively represents any $\varepsilon\in\mathcal{T},$  $\Gamma_{\varepsilon}(\lambda,\lambda')\ge0\;\; \forall \lambda,\lambda'\in\Lambda.$ Given a state $\rho\in\mathcal{S}$ which is represented by the non-negative quasiprobability distribution $\mu_{\rho}(\lambda),$ the state $\rho'=\varepsilon(\rho)$ is also non-negatively represented, as $\mu_{\rho'}(\lambda)=\sum_{\lambda'}\mu_{\rho'}(\lambda')\Gamma_{\varepsilon}(\lambda,\lambda').$ This proves the first part of the proposition.
The converse implication does not hold. In the example of the single qubit stabilizer theory all the transformations map non-negative states to non-negative states (Wootters-Gibbons' Wigner function \cite{Gibbons2004,Cormick2006}), but the theory, with positivity preserving transformations and non-negative states and measurement elements, does not allow for a transformation noncontextual model.
The core reason why transformation contextuality, and, equivalently, the unavoidable presence of some negativity in $\Gamma_{\varepsilon},$ is a weaker notion of nonclassicality than the breaking of positivity preservation is that $\Gamma_{\varepsilon},$ despite assuming some negative values, can still preserve the positivity between the quasiprobability representations of the states.  
\end{proof}

\begin{figure}[h!]
\centering
{\includegraphics[width=.65\textwidth,height=.15\textheight]{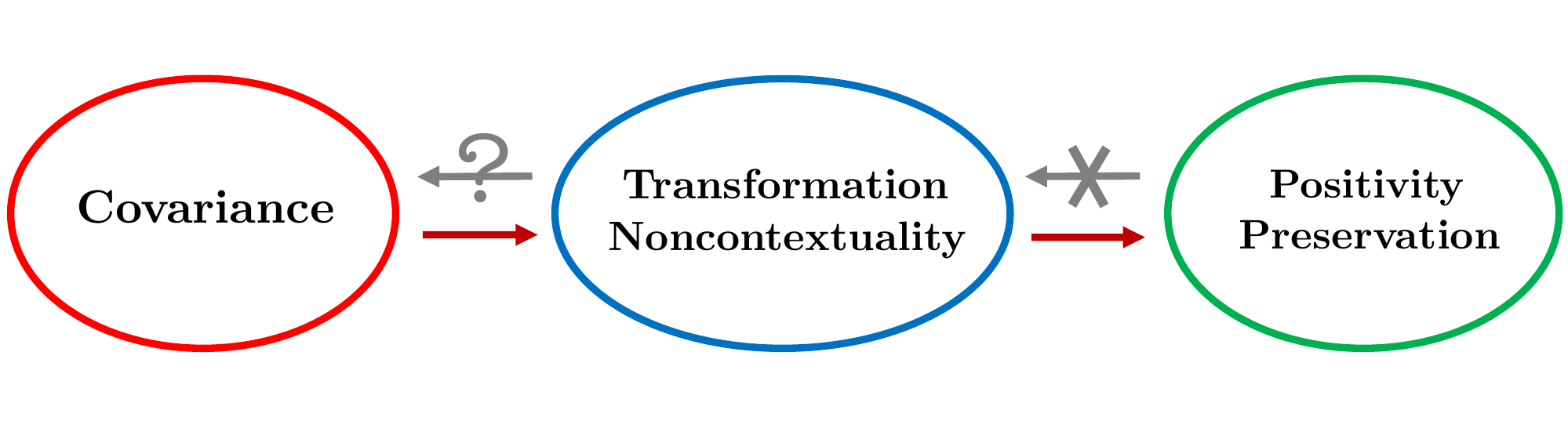}}
\caption{\textbf{Relationship between three notions of classicality associated to quantum transformations.} Given a subtheory of quantum theory $(\mathcal{S},\mathcal{T},\mathcal{M})$, where the set of transformations $\mathcal{T}$ is defined specifying a group of unitaries and contains transformations that map between states in $\mathcal{S}$ (and measurements in $\mathcal{M}$), the existence of a Wigner function that represents it covariantly implies the existence of a transformation noncontextual ontological model for it (Proposition \ref{CovarianceTransfNC}). Regarding the converse implication, we have conjectured that the existence of a noncontextual ontological model for the subtheory implies covariance if the model is the one provided by the Wigner representation (Conjecture \ref{TransfNCCovariance}). Both covariance and transformation noncontextuality imply the existence of a positivity preserving quasiprobability distribution for the subtheory, but not vice versa (Propositions \ref{TheoremCovariance} and \ref{TheoremTransfNC}). 
}
\label{Relations}
\end{figure}

The relations found are depicted in figure \ref{Relations}.

\onecolumngrid

\twocolumngrid


\section{Discussion}
\label{Discussion}


The results obtained regard the relationships between a property of Wigner functions -- covariance, a property of ontological models -- transformation noncontextuality, and a property of quasiprobability representations -- positivity preservation. It would be interesting to extend the notion of covariance to more general quasiprobability representations. However, we argue that a relation between this extended covariance and transformation noncontextuality is not expected to hold. Such covariance would hold in the CSS rebit subtheory studied in \cite{Delfosse2015}, while transformation noncontextuality is violated\footnote{The two decompositions, $\rho \rightarrow 1/2(\rho + Y\rho Y)$ and $\rho \rightarrow 1/2(X\rho X + Z\rho Z),$ correspond to the same channel in the CSS subtheory of \cite{Delfosse2015}, even if any ontological model of the theory represents them as distinct (the proof mirrors the one contained in \cite{Lillystone2018} for the single qubit stabilizer theory). The author thanks Piers Lillystone for making him aware of this fact.}. 
Still, the latter is not expected to imply covariance, considering that the orthonormality property of the phase point operators -- crucial for proving Proposition \ref{CovarianceTransfNC} -- does not hold for generic quasiprobability representations.

One extra way to generalize covariance would be to consider subtheories with arbitrary CPTP maps, where any CPTP map $\varepsilon$ is assumed to be given by a unitary $U$ acting on a Hilbert space that describes both the system and the environment, $\varepsilon(\rho)=\sum_k E_k\rho E_k^{\dagger}=\textrm{Tr}_E[U\rho_{SE}U^{\dagger}],$ where $\rho_{SE}$ is the state of the system and environment, while $\rho$ is the state of the system only. Covariance (see equation~\ref{covarianceA}) is then defined as $UA_{SE}(\lambda)U^{\dagger}=A_{SE}(S_{SE}\lambda+a_{SE}),$ where $S_{SE}$ and $a_{SE}$ are a symplectic matrix and a vector acting on the phase space associated to the system and environment, for all the possible unitaries that define $\varepsilon.$ 
We leave the study of this notion for future research. 

The main open question about this work regards Conjecture \ref{TransfNCCovariance}. A possible strategy to prove it would be to first leverage on the ``proof of the argument in support of the conjecture'' -- showing that the existence of a transformation noncontextual ontological model corresponding to a Wigner representation of the subtheory $(\mathcal{S},\mathcal{T},\mathcal{M})$ implies that phase point operators are mapped between themselves by the unitary transformations in $\mathcal{T}$. Then, one may argue this to imply that the unitaries in $\mathcal{T}$ must be Clifford unitaries, as a non-Clifford transformation could not map phase point operators (that are sum of Pauli operators) among themselves. Finally, the diffucult part would be to prove, for these unitaries, a result akin to Lemma 9 (and then 11) in \cite{Delfosse2015}. 
Notice that, if Conjecture \ref{TransfNCCovariance} holds true, then we can restate the results of Proposition \ref{CovarianceTransfNC} and Conjecture \ref{TransfNCCovariance} by saying that the set of transformations that are positively represented by a given Wigner function contains all and only those transformations that are represented covariantly by the Wigner function.

Another open question is whether any quasiprobability representation on an overcomplete frame \cite{FerrieEmerson2009} (like the one in \cite{Delfosse2015} and unlike the Wigner representation) always implies the corresponding model to be transformation contextual.  In an overcomplete frame representation the phase-point operators are, by definition, more than the ones needed to form a basis. This is expected to imply multiple distinct representations of each transformation.\footnote{By the time this article first appeared this result has been proven in \cite{SchmidPusey2020} with minor extra assumptions.} 
Showing that only complete frame representations provide noncontextual models would be a first step to ultimately try to prove that any noncontextual ontological model has to correspond to a Wigner representation. 

Our results show that breaking positivity preservation is a stronger notion of nonclassicality than transformation contextuality and non-covariance. With respect to positivity preservation, the single qubit stabilizer theory is classical, even if it shows transformation contextuality and breaks covariance. This fact motivates the study of the physical justifications for considering positivity preservation a legitimate classical feature. It would be also interesting to explore other subtheories in which positivity preservation, unlike other notions of classicality, coincides with classical computational simulability. With this work we aim to promote further research on the reasons why the contextuality present in qubit stabilizer theory has, computationally, a classical nature.\footnote{Meaningful results in this direction have been found in \cite{Karanjai2018}, where it is shown that contextuality lower bounds the number of classical bits of memory required to simulate subtheories of quantum theory. In the case of multi-qubit stabilizer theory, it demands a quadrating scaling in the number of qubits, which coincides with the scaling of Gottesman-Knill algorithm \cite{Gottesman99}. If one considers the noncontextual stabilizer theory of odd dimensional qudits the scaling becomes linear.}

\section*{Acknowledgements}

The author thanks Huangjun Zhu for contributing in proving the second part of the argument in support of conjecture 1, and David Schmid and Matt Leifer for useful discussions. This research was supported by the Fetzer Franklin Fund of the John E. Fetzer Memorial Trust while at Chapman university and by the Einstein Research Unit `Perspectives of a Quantum Digital Transformation' while at Technische Universit\"{a}t Berlin.

\bibliographystyle{apsrev4-1}
\bibliography{Covariance_TransfNC.bib}

\end{document}